\documentstyle[12pt]{article}
\newcommand{\amp}{{\cal A}}
\newcommand{\kta}{{\vec k_a}}
\newcommand{\ktb}{{\vec k_b}}
\newcommand{\ktc}{{\vec k_c}}
\newcommand{\q}{\Delta}
\newcommand{\qt}{\vec\Delta}
\newcommand{\nnn}{\noindent}
\catcode`\@=11

\@addtoreset{equation}{section}

\begin{document}
\baselineskip .25in
\renewcommand{\thefootnote}{\dag}
\newcommand{\numero}{McGill 93-25}
\newcommand{\numero2}{hep-ph/9310298}

\newcommand{\auteura}{J.R. Cudell\footnote[1]{cudell@hep.physics.mcgill.ca}
\renewcommand{\thefootnote}{\ddag}
 and B.U. Nguyen\footnote[2]{bao@hep.physics.mcgill.ca}}
 \newcommand{\auteurb}{Department of Physics\\
McGill University\\ 3600 University Street\\
Montr\'eal, Qu\'ebec, H3A 2T8, Canada}
\newcommand{\beq}{\begin{equation}}
\newcommand{\beqn} {\begin{eqnarray}}
\newcommand{\r}{{\vec r}}
\newcommand{\eeq}{\end{equation}}
\newcommand{\eeqn} {\end{eqnarray}}
\newcommand{\pa}{{p}^a}
\newcommand{\pbb}{{p}^b}
\newcommand{\titre}{A consistent next-to-leading-order QCD calculation \\
of hadronic diffractive scattering }

\newcommand{\abstrait}{We calculate the order $\alpha_s^2$ and order
$\alpha_s^3$ QCD contributions to colour-singlet exchange
in the leading $\log s$ approximation. We implement the resulting amplitude
at the hadronic level and thus construct the QCD pomeron
and odderon to this order of perturbation theory.
We show that the structure of the hadronic form factors provides
a natural mechanism through which the odderon gets suppressed at
$t=0$ whereas it dominates the elastic cross section at large $t$.
We also demonstrate that the inclusion of nonperturbative effects through a
modification of the gluon propagator accelerates greatly the
convergence of the $\log s$ expansion, although not enough to provide
agreement with the data.}
\begin{titlepage}
\hfill \numero \\
\hspace*{\fill}{\numero2}\\
\vspace{.5in}
\begin{center}
{\large{\bf \titre }}
\bigskip \\ by \bigskip \\ \auteura \bigskip \\
\renewcommand{\thefootnote}{\ddag}
 \bigskip \auteurb
 \bigskip \\

\renewcommand{\thefootnote}{\dag }
\vspace{.9 in}
{\bf Abstract}
\end{center}
\abstrait
 \bigskip \\
\end{titlepage}

\section{Introduction}

Hadronic diffractive scattering has defied theoretical understanding
for the past twenty years. Although the data can be fitted remarkably well
to a form that was predicted by the old Regge theory \cite{LD}, perturbative
QCD does not give rise to a pomeron and an odderon which agree with
data. What perturbative QCD gives us instead is the so-called ``perturbative
pomeron'', which seems to be a theoretical construct in need of experimental
confirmation.

This ``perturbative pomeron'' results from the formidable machinery of the
Balitski\v \i-Fadin-Kuraev-Lipatov (BFKL) equation \cite{BFKL}, and
has the following drawbacks:\\
$\bullet$ Its leading contribution to the hadronic amplitude, for fixed
$\alpha_s$ goes like $s^{1+12\log (2)\alpha_s/\pi}$, {\it i.e.} for any
reasonable value of $\alpha_s$,
it increases much faster than the data, which behave like
$s^{1.08}$ \cite{LD}.\\
$\bullet$ At nonzero $t$, the differential elastic cross section
has the wrong shape:
its logarithmic slope at $t=0$ is infinite and its curvature is too big. \\
$\bullet$ A more subtle problem stems from the fact that the pomeron-hadron
vertex is almost local, {\it i.e.} similar to a photon exchange term, with the
spin structure $\gamma^\mu$. This property is
necessary to reproduce the factorizability of the pomeron amplitudes
\cite{goulianos}.
QCD exchanges do not reduce to an effective photon-like
interaction. The quark-quark interaction does, but the problem comes from the
implementation of quark-quark scattering inside a hadron. Quark-quark
scattering is infrared divergent. However, we know that colour-singlet objects
do not interact at infinite distances. So the wavefunction effects have to
cancel the infrared divergence. Unfortunately, the side-effect of this
cancellation is to destroy the factorizability of a simple
$\gamma^\mu$ vertex \cite{Richards}.

Another perturbative construct is the odderon. Although we shall not be able
to evaluate its intercept at this order, we shall be able to tackle the
following problem: experimentally, the ``Landshoff term'' \cite{Lterm},
which arises from 3-gluon exchange, each gluon being attached to a
different quark, gives the leading contribution
to the elastic cross section at large $t$. However, we know that at small $t$,
this odderon contribution is small. We shall explain how this feature emerges
naturally for the lowest-order odderon.

Thus the three main questions we want to answer are the following:
how can we reduce the
intercept of the QCD pomeron? how can we limit the effects of the wave
function? Why is the odderon negligible at small $t$ and dominant at high $t$?
Rather than trying to settle these issues through the BFKL
equation \cite{RH}, we shall perform a complete third-order calculation
(which is
the first order at which one can see the pomeron intercept and slope) of
hadron-hadron scattering. We
shall show that all the problems of the leading-log resummation are present at
this order of perturbation theory, and we shall think of it as a first-order
Taylor expansion of $As^{\alpha_0+\alpha ' t}$ in $\log s$. We can then
directly use it as a laboratory in which we can test some of the
ideas that have been proposed to resolve these puzzles.

The first class of ideas is to limit the wave function effects by assuming
that the dynamics of the QCD vacuum provides an intrinsic infrared regulator.
This idea was proposed by Landshoff and Nachtmann
\cite{LN},
who used an {\it ad hoc} form for the gluon propagator that enabled them to
perform the required integrals. Since then, the infrared properties of the
gluon propagator have been re-evaluated from a more theoretical point
of view: solutions to the Dyson-Schwinger
equation for the gluon propagator have indeed been found, solutions
which are smoother than a
pole in the infrared region \cite{axial,feynman}.
Also, some earlier work has proposed that
the gluon develops a mass at zero momentum transfer \cite{Cornwall}.
We shall show here that the perturbative
calculation can be easily recast in a form which
allows for a modification of the gluon propagator, provided that there exists
a K\"allen-Lehmann density: the complete third-order result then
depends only on the $t$-channel propagators.

The second possibility is to study the form factors more closely. The
calculation will involve two form factors in the pomeron case, and an extra
one in the odderon case. One can either make a model for those, which will
depend on an explicit proton wavefunction, or assume a simple
parametrization. We shall pursue both routes.

The paper is organized as follows: in the second section, we re-examine the
two-gluon calculation, set up most of the formalism and discuss the use of
nonperturbative propagators at this order.
The third section details the calculation of the amplitude to
order $\alpha_s^3$, first in the odderon case, then in the pomeron case.
The fourth section is devoted to a discussion of our results.
Note that we shall work in the Feynman gauge throughout.

\section{Two-gluon exchange}

Since the work of Low and Nussinov \cite{Low}, it has been realized that the
lowest-order QCD exchange that has the quantum numbers of the pomeron (charge
parity $C=+1$, colour singlet) is two-gluon exchange.

As QCD is a theory of quarks and gluons, we must first consider quark-quark
scattering. The lowest-order diagrams contributing to colour-singlet exchange
are shown in Figure~1. The two incoming quarks have momenta $\pa$
and $\pbb$, which, in the limit $s\approx2\ \pa\cdot\pbb\rightarrow \infty$,
can be
identified with the + and $-$ directions of light-cone kinematics
($p_+=p_0+p_3$ and $p_-=p_0-p_3$). If we
consider elastic scattering, both the initial-state and the final-state quarks
have to be near shell.
That means that the momentum transfer
$t=\q^2$ must be in the transverse direction, $\q\approx (0,\qt,0)$
so that both $(\pa+\q)^2$ and $(\pbb-\q)^2$ remain small as
$s\rightarrow \infty$.
(From now on, we shall denote the transverse direction by an arrow.)

In two-gluon exchange, the momentum kick $\q$ is split into two gluons
$\q=k^a+k^b$. The diagrams of Figure~1 involve integrals of the form:
\begin{equation}
\int dk^a_+ dk^a_- d\kta {s^2\over
[(p^a-k^a)^2-m^2]\ [(p^b+k^a)^2-m^2]\ [(k^a)^2]\ [(\q-k^a)^2]}\label{qqscatt}
\end{equation}
The leading contribution comes from the region when one (and
only one) of the quark propagators is of the order of ${s}$. This
happens if the gluons carry a large component either in the + or in the
$-$
direction. The amplitude then contains a factor {\it e.g.} $\int {dk^a_-
/(p^a_+ k^a_-)}\approx
{(\log s)/ s}$. However, as $u\approx -s$, crossing symmetry implies that
one graph behaves as $s \log s$ and the other as $-s \log(-s)$. Thus the
two real parts cancel, leaving only an imaginary part.

In order to calculate this imaginary part, we can use cutting
rules: in graph~1(a) we put the intermediate quarks on-shell, replacing
their propagators by a delta function of their 4-momentum squared.
The $k^a_-$ and $k^a_+$ integrals can
then be done trivially, and bring a factor $1/s$. The same argument as for
$\q$ holds, and the gluons have to be transverse:
$k^a\approx\kta$, $k^b\approx \ktb$.

To sum up, two-gluon exchange takes the following form:
 \beq
\amp_2^q(s,t)=i \alpha_s^2 s C_2 \int d\kta d\ktb
{d\amp_2^q\over d\kta d\ktb}\label{qqscatt2}
\eeq
with
\beq
{d\amp_2^q\over d\kta d\ktb}=
\delta^{(2)}(\qt - \kta - \ktb) {1\over (\kta^2+\sigma_a)}\times{1\over
(\ktb^2+\sigma_b)}\label{qqscatt3}\end{equation}
where we have introduced two gluon squared masses $\sigma_a$ and $\sigma_b$,
which for now can be considered as infrared regulators, and a colour factor
$C_2=8/9$.

\subsection{Photon-photon scattering}

Quark-quark scattering lacks a crucial property of hadron-hadron scattering:
as their wavelength increases, one expects that gluon interactions will average
out the colour of the hadron, and thus effectively decouple. In
other words, the
calculation should be infrared finite. To verify this fact, we need a model
for a hadron.

The simplest object for which the above argument holds is a photon,
and since the work
of Cheng and Wu, and Frolov, Gribov and Lipatov \cite{photon},
it is known that the
photon-photon scattering amplitude takes the form:
\beq
{\cal A}_2 \sim
i s \int d\kta d\ktb \ {d\amp_2^q\over d\kta d\ktb}
 [{\cal E}_1^\gamma(\qt^2)-{\cal E}_2^\gamma(\kta,\ktb)]^2\label{gagascatt}
\eeq
${\cal E}_1^\gamma$ comes from graphs where the gluons
are attached to the same quark line, see Figure~2(a),
and takes the following form:
\beq
{\cal E}_1^\gamma(\qt^2)= 4\alpha_{em}
\int_0^1 d\beta\int d\vec p \ \psi^*(\beta,\vec p+{\beta\qt\over 2})
 \psi(\beta,\vec p-{\beta\qt\over 2}) \label{f1gg}\eeq
$\psi(\beta,\vec p)$ is the wavefunction of a quark produced out of the photon
with longitudinal momentum fraction $\beta$ and transverse momentum $\vec p$.
It
can be shown to be:
\beq
\psi(\beta,\vec p)\sim {\sqrt{\beta(1-\beta)}\over \vec p^2} {\cal
P}_{ij}(\vec p,s)\label{psigg}\eeq
with $ {\cal P}_{ij}(\vec p,s)$ a factor that depends on the polarizations
of the incoming and outgoing photons.
${\cal E}_2^\gamma$  comes from diagrams
where the gluons are attached to different quark
lines. For instance, Figure~2(b) will produce a
term ${\cal E}_1^\gamma{\cal E}_2^\gamma$ in
Eq.~(\ref{gagascatt}).
This form factor
can be written:
\beq {\cal E}_2^\gamma(\kta,\ktb)={\cal E}_1^\gamma
\left((\ktb(1/\beta-1)-\kta)^2\right)\label{f2gg}\eeq

The important lesson is that
$[{\cal E}_1(\qt^2)-{\cal E}_2(\kta,\ktb)]^2\sim \kta^2$ as
$\kta\rightarrow 0$ (and similarly for $\ktb$), {\it i.e.} the infrared
divergence
of the gluon propagator is cancelled by the form factor. Also, the
particular form taken by ${\cal E}_2^\gamma$ comes from
the symmetry of the vertex
$\gamma\rightarrow q\bar q$, which dictates for instance that
$\psi(\beta,\vec p)=\psi(1-\beta,\vec p)$. This property must be conserved in
the pion case, but need not be present in the baryon case.

\subsection{Hadron-hadron scattering}

The preceeding formalism finds a natural extension in the work of Gunion and
Soper \cite{GS}. Here and in the following, we shall assume that hadrons are
composed of
their valence quarks only. This amounts to saying that sea partons are
generated by higher-order corrections.

In the high-$s$ limit, the two incoming hadrons are living on the light-cone,
{\it i.e.} at fixed $x_-$ or $x_+$. These two directions then alternatively
play the role of
time in the definition of the wavefunctions $\psi$. Transforming the remaining
free $+$ or $-$
components to momentum space, hadrons are thus described by a wavefunction
$\psi(\{\beta_j\},\{\r_j\})$, with $\beta_j$ the longitudinal momentum fraction
of quark $j$ and $\r_j$ its impact parameter in the center-of-momentum frame of
the hadron. This wavefunction is a priori unknown.
One can then show, in the eikonal approximation, that multi-gluon exchange
takes the following form \cite{GS}:
\beqn
\amp_{\infty}&=&-2 i s \int d{\vec b} e^{-i\qt.{\vec b}} \nonumber\\
&\times & [\prod_{j=1,n_q} \int^{1}_{0} d\beta'_j \int d{\vec r}_j]
|\psi(\beta'_j,\r_j)|^2
[\prod_{l=1,n_q} \int^{1}_{0} d\beta_l \int d{\vec s}_l]
|\psi(\beta_l,{\vec s}_l)|^2 \nonumber\\
&\times & \left( exp \left( - i {2\pi\alpha_s} \sum_{i,j} \sum_{c=1,8}
\lambda^{c}_{i} \lambda^{c}_{j} V({\vec x}_i-{\vec x}_j) \right) - 1
\right)
\label{eikonal}\eeqn
with
\beq
V({\vec x}) = - \int dx^{+} dx^{-} \Delta_{F} (x^{+},x^{-},{\vec x})
\label{gluprop}
\eeq
where one implicitly assumes an ordering of the vertices in the $+$ and $-$
directions. The key issue for this eikonal formula to be valid is that the
scattering of the quarks lying at fixed $p^-$ (respectively $p^+$)
must not give final state quarks with a substantial $+$
(resp. -) component, so that they can recombine into hadrons. More precisely,
this ``opposite'' component must be of the order of $1/s$. As we shall see
later, large rapidity gaps imply that some of the diagrams are
suppressed for this reason, and we will have to subtract their contribution
from expression (\ref{eikonal}).
As from the argument leading to Eq.~(\ref{qqscatt2}),
large rapidity gaps
do not contribute to two-gluon exchange, we shall not pursue the matter further
here.

The eikonal formula for the scattering of two hadrons $h_1$ and $h_2$
containing $n_1$ and
$n_2$ valence quarks via the exchange of two gluons is obtained by expanding
Eq.~(\ref{eikonal}) to order $\alpha_s^2$. This gives:
\beq
\amp_2=i\alpha_s^2s C_2 n_1 n_2\int d\kta d\ktb {d\amp^q_2\over d\kta d\ktb}
[{\cal E}_1^{h_1}(\kta+\ktb)-{\cal E}_2^{h_1}(\kta,\ktb)]
[{\cal E}_1^{h_2}(\kta+\ktb)-{\cal E}_2^{h_2}(\kta,\ktb)]
\label{hhscatt}\eeq

\beq {\cal E}_1(\qt)=\int d{\cal M} e^{i \qt.\r_k}\label{f1}\eeq
and
\beq {\cal E}_2(\kta,\ktb)=\int d{\cal M} e^{i
\kta.\r_k+i\ktb.\r_l}\label{f2}\eeq
with $l\neq k$.
The natural integration measure $d{\cal M}$ is defined as:
\beq d{\cal M}=[\prod_{j=1,n_q} d\beta_j d\r_j ]\delta^{(2)}
(\sum_{j}\beta_j \r_j)
\delta(\sum_{j}\beta_j-1) |\psi(\beta_j,\r_j)|^2 \label{measure}\eeq
The first delta function defines the center of momentum of the hadron, whereas
the second one enforces longitudinal momentum conservation.
Assuming that hadrons are made of valence quarks only, we normalize the
wavefunction according to:
\beq \int d{\cal M}=1 \label{norm}\eeq

The same model applied to $\gamma$p elastic scattering leads to the
identification of ${\cal E}_1$ with the Dirac
elastic form factor $F_1$. ${\cal E}_1$ is thus
measured directly so that
we know its form from experiment. In the proton case, we shall use a dipole
electric form factor $G_E=(1-t/0.71)^{-2}$, which
fits the data to $t=10$~GeV$^2$ at least
\cite{ppform}.
This leads to the Dirac form factor:
\beq
{\cal E}_1(t)={(3.53-2.79 t)\over(3.53-t)\ (1-t/0.71)^2}
\label{f1pp}
\eeq
In the pion case, the form factor has been
measured only to moderate values of $t<0.3$~GeV$^2$,
so that we have to assume the functional form:
\beq
{\cal E}_1(t) = {1\over (1-t\ <r_\pi^2>/6) } \label{f1ppi}
\eeq
where $\sqrt{<r_\pi^2>}=0.663$ fm is the
electromagnetic radius of the pion \cite{piform}.

The form of ${\cal E}_2$ is more arbitrary
as the only firm property that can be
established from Eq.~(\ref{f2}) is the cancellation of the infrared divergences
as $\kta$ or $\ktb\rightarrow 0$, {\it i.e.} ${\cal E}_1\rightarrow
{\cal E}_2$.
For the purpose of the present
analysis, we adopt two strategies. The first is to allow ${\cal E}_2$ to vary
according to a functional form guaranteeing infrared finiteness:
\beq
{\cal E}_2(\kta,\ktb)={\cal E}_1(\kta^2+\ktb^2-f\kta.\ktb)\label{f2simple}
\eeq
The appropriate values in the pion case is $f=2$ and in the proton case
$f=1$, corresponding to a wavefunction peaked at
$\beta=1/2$, and $\beta=1/3$ respectively.
The fact that
$f=2$ in the pion case is obvious: if both quarks are deflected by an identical
amount in the transverse direction, and if their longitudinal momenta are
equal, then they must recombine into a pion:
\beq
{\cal E}_2^\pi(\vec k,\vec k)=1
\label{Ltermpi}
\eeq
The parametrization (\ref{f2simple}) of the pion form factor,
together with the value $f=2$, result
from the symmetry property (\ref{f2gg}), transposed to the case of a pion
at fixed $\beta=1/2$.
We shall see in section 3.2 why $f$ must be equal to 1 in the proton case
\footnote{In refs. \cite{axial,halzen}, the
value $f=7$ was erroneously used for proton-proton scattering.}.

The second strategy is to choose an explicit form for the hadron wavefunction.
Assuming that the longitudinal momentum is distributed as in the valence quark
structure function, we use:
\beq
\psi_{abc}(\{\vec r_i\},\{\beta_i\})=
N \epsilon_{abc}/\sqrt{6}\left(\prod_{i=1,3} {(1-\beta_i)^{n}\over
{\sqrt \beta_i} }
\right) \times \exp \left( {\sum_i \r_i{\ ^2}\over r_h^2}\right)
\label{psih}
\eeq
with $N$ a normalization factor, $n=3$ for protons and 1 for pions,
$r_h$ a hadronic radius and $abc$ the colour of the quarks.

As can be seen from Figure~3, this form reproduces well the parametrization
(\ref{f1pp}) for $r_p=0.65\  fm$ while producing a
physical model for ${\cal E}_2$.
Note that for moderate values of $|\kta|,\ |\ktb|$ the effect of the smearing
of $\beta$ is to shift $f$ in Eq.~(\ref{f2simple}),
to $f\approx 0.8$ in the proton case and $f\approx 1.5$ in the pion case.

We give in Table 1 the results of perturbative two-gluon exchange.
We have considered the
calculation either at fixed $\alpha_s$, or for a running
$\alpha_s(q^2)=(12\pi/ 27)\log(q^2/\Lambda_{QCD}^2)$,
which we freeze once it gets to $\alpha_s=1$. (We have
taken $\Lambda_{QCD}^{(3)}=0.3$~GeV, and assumed that the scale of $\alpha_s$
is the virtuality of the gluon attached at the vertex.)
We also show the dependence of some of
the results on our choice of form factor:
the numbers in parenthesis are obtained
using the model (\ref{psih}) for the proton
wavefunction, and the others correspond to
the simple parametrization (\ref{f1pp}-
\ref{f2simple}), both in the case of a fixed and of a running $\alpha_s$.

Two-gluon exchange sets the scale correctly: the $pp$ cross section  is about
$76\alpha_s^2$ mb, which for $\alpha_s\sim 0.5-1$ is of the right order of
magnitude. Furthermore, the quark counting rule
can be reproduced only if we assume that
the ${\cal E}_1-{\cal E}_2$ factors are not
too different when going from the proton to the
pion. If we use the form (\ref{f2simple}) with f=2 for pions
and f=1 for protons, it turns out that
the ratio $\sigma_{p\pi}/\sigma_{pp}=0.65$ is close
to the experimental number~0.62. Let us point
out however that this is very sensitive to
our input value for $<r_\pi^2>$, and
to the assumed functional dependence of Eq.~(\ref{f1ppi}).
In perturbative QCD, the quark counting rule
would then be only an accident, and the factorizability of the pomeron exchange
\cite{goulianos} would be lost. However, the main
problem comes from the elastic differential cross section.
Its shape, as shown in Figure~4,
comes out wrong: instead of an exponential, it has too much
curvature, and its logarithmic slope at the origin turns out to be infinite.

These problems can be traced to the infrared region. To make this point clear,
we show in Figure~5 the average momentum
$k_{ave}=<|\kta|>$ in formula (\ref{hhscatt})
as a function of $t$. We see that it
increases with $-t$, from a fraction of a~GeV
near $t=0$ to about 1.5~GeV near $t=-10$~GeV$^2$. This means that even at high
$-t$ the small-$k^2$ region will dominate: the truly perturbative region plays
a negligible role in diffractive scattering. This feature is identical at
order $\alpha_s^3$. It is thus important to examine
this region in detail, and to see how the most modest changes in the
infrared region can affect our results.

 \subsection{Nonpertubative effects}

Landshoff and Nachtmann \cite{LN} have argued that a modification of the gluon
propagator can produce a factorizing amplitude, which in turn should remove
the
wavefunction dependence, restore factorization, and make $B(0)$ finite.

In order to use a modified gluon propagator, we must first observe that the
infrared regulators of formulae (\ref{qqscatt2}) can be treated as the
squared masses that enter the
K\"allen-Lehmann representation for the propagator:
\beq
D(q^2)=\int_0^\infty d\sigma {\rho(\sigma)\over q^2-\sigma+i\epsilon}
\label{KLehmann}
\eeq
Note that in fact we do not need a
K\"allen-Lehmann density {\it per se}, but simply the existence
of a Hilbert transform, {i.e.} neither the range of integration nor the
reality of $\rho$ will matter in the following.

One can then repeat the perturbative calculation by commuting the $\sigma$
integrals and the $d^4 k$ integrals. One obtains the same results
(\ref{qqscatt2}, \ref{qqscatt3}, \ref{gagascatt}),
which then need to be convoluted with the
K\"allen-Lehmann densities $\rho(\sigma_1)$, $\rho(\sigma_2)$. This
reconstructs the propagators, and the ${1\over k^2+\sigma}$ then becomes
$-D(-k^2)$ because of Eq.~(\ref{KLehmann}).

We shall see that this property of the amplitude is encountered again at the
next order of perturbation theory: the perturbative propagators can be
replaced by nonperturbative ones, and provided that there exists a Hilbert
transform, we can justify this substitution on theoretical grounds.

One might however still worry about the gauge invariance of the theory: if
the structure of the QCD vacuum changes the
propagators, then in order to obey the Ward-Slavnov-Taylor (WST) identities,
the vertices need to be modified, and probably
the quark propagators, too. As far as the vertices go, one can argue that if
the K\"allen-Lehmann density $\rho(\sigma)$
is concentrated near the origin, then the corrections to
the vertices are probably of the order of $\sigma$, and thus
these contributions are going to be small (this is confirmed by the results of
Ref.~\cite{BFKL}, where gauge invariance is maintained via a Higgs mechanism).
As far as the quarks are concerned, we
simply assume that confinement affects the gluons at shorter distances than
it affects the quarks. This does not break the gauge invariance of the
$O(\alpha_s^2)$ calculation, as the gluon propagator does not enter the
quark-gluon WST identities. Recent results on the quark propagator in
the axial gauge confirm that a simple pole in the quark propagator
is not far from the true solution \cite{CR}.

Hence the replacement of the
gluon propagator by a nonperturbative counterpart does not violate gauge
invariance for gluon exchange diagrams. One can think of this replacement as
the inclusion of a subclass of diagrams (the gluon self-energy) which are
resummed via nonperturbative methods (the DS equations). Although these
diagrams are supposedly sub-leading $\log s$, their inclusion certainly
changes the leading-log answer.

The contributions to proton-proton scattering at $t=0$ are of two kinds: the
${\cal E}_1^2$ term of formula (\ref{hhscatt}) leads
naturally to the quark counting
rule and to an effective pomeron vertex behaving like
$\gamma^\mu$ \cite{LN,book}. This is due to the fact that along two quark
lines of momentum $p_1$ and $p_2$ we can write $[\gamma_\mu (p_1.\gamma)
\gamma_\nu]\otimes[\gamma^\mu (p_2.\gamma) \gamma^\nu]\approx 2
s\gamma_\mu\otimes\gamma^\mu$. On the other
hand, the terms containing ${\cal E}_2$
do not have such a simple structure, as they involve the hadronic
wavefunction in the middle of the preceeding expressions.

However, this hadronic wavefunction must contain a scale, the hadronic radius
$R$.
If gluons propagate only to finite distances,
{\it i.e.} if the gluon propagator is smoothed in the infrared
region, then one can produce a damping of the ${\cal E}_2$ terms. A simple
dimensional argument leads to the conclusion that a damping in the
infrared region means the introduction of a nonperturbative scale $\mu_0$,
such that the gluon propagator now looks like:
\beq
D(q^2)={1\over\mu_0^2}\  {\cal D}({q^2\over\mu_0^2})
\label{npglu}
\eeq
with $\cal D$ a function without a pole at the origin.
In the case $\mu_0>>1/R$, the propagator hardly changes while the form factor
drops sharply, so that one gets a suppression factor $(\mu_0 R)^{-2}$. This
suppression of the ${\cal E}_2$ terms in turn leads to a better agreement of
the elastic differential cross section with experiment, as it is already known
that a $\gamma^\mu\otimes\gamma_{\mu}$ combined with the parametrization
(\ref{f1pp}) leads to a good fit to the data \cite{LD}.

The problem remains however to see whether such a simple idea can be
theoretically justified. One indeed expects gluons not to propagate to
infinity: their self-interaction should confine them, and thus modify
their propagator.
Lattice studies suggest that the gluon
propagator is suppressed at small momentum
transfer \cite{lattice}, but its exact
structure remains unclear.
By studying the Dyson-Schwinger equation for the gluon
propagator \cite{axial,feynman} or
the Gribov horizon \cite{Zwanziger}, several
groups have recently confirmed that such a picture is essentially correct.
the Dyson-Schwinger equations have been shown to possess, in their truncated
form, at least two types of solutions: those that behave like $1/k^4$ at the
origin, as well as smoother ones. As the $1/k^4$ solutions do not have a
Hilbert transform, we do not know how to use
them in diffractive calculations. We thus
assume that the other solutions, which do not have poles for $q^2\leq 0$,
are those which are relevant for diffractive scattering.
These solutions have been derived in various gauges.
In the following, we shall consider only
gauge-invariant sets of diagrams, so that the dependence on the propagators
comes from the different approximations made by the authors of references
\cite{axial,feynman}.

Although the
asymptotic forms of \cite{axial,feynman}
agree with perturbative QCD at large $k^2$, there is a wide
disagreement as to the details of their
behaviour near $k^2=0$: they either go to
zero \cite{feynman,Zwanziger}, or are finite \cite{Cornwall}, or are infinite
\cite{axial}. We limit ourselves here to the
study of three propagators which represent the whole range of behaviours at
the origin which can be implemented in the calculation (assuming that the
propagator does not change sign for $q^2<0$.) Note that we have
considered several other possibilities (massive propagators, exponential
propagators, perturbative propagator with a cutoff) with similar results
\cite{bao}.

The H\"abel-K\"onnig-Reusch-Stingl-Wigard (HKRSW) propagator \cite{feynman}
vanishes at the origin.
Its form has been suggested by a
consistency argument in the Landau gauge, and agrees with that derived by
Zwanziger based on considerations related to the Gribov horizon
\cite{Zwanziger}.
\begin{equation}
D(k^{2})= {1\over  k^{\ 2} + \mu_0^4/k^{2} }\label{HKRSW}
\end{equation}

In the axial gauge, Cornwall has derived a gauge-invariant set of diagrams
defining a gluon mass \cite{Cornwall}.
Although one might worry about simply putting this mass
into an explicitly gauge dependent object, it enables us to consider the
possibility of a propagator finite at the origin:
\begin{equation}
\alpha_s D(k^{2})= \left\{{27\over 12\pi} \ (k^{\ 2} + \mu^2(k^2) )\ \log\
\left( {(k^{2} + \mu^2(k^2))\over \Lambda_{QCD}^2}
\right)\right\}^{-1}\label{Cl}
\end{equation}
\noindent where:
\begin{equation}
\mu^2(k^2)=\mu_0^2 \left( {\log (4\mu_0^2/\Lambda_{QCD}^2) \over
\log ((k^{2}+ 4\mu_0^2)/\Lambda_{QCD}^2) }
\right)^{12/11}\label{Cornwalle}
\end{equation}

Notice that in the axial gauge, a theorem due
to Baker, Ball and Zachariasen implies that the gauge-dependent
propagator is infinite at the
origin \cite{BBZ}. Solutions have been found
by D.A. Ross and one of us \cite{axial} which behave
like a fractional power of $k^2$ near $k^2=0$:
\begin{equation}
D(k^{2})={1\over \mu_0^2\left( 0.88\left({k^{2}\over
\mu_0^2}\right)^{0.22}
-0.95\left({k^{2}\over \mu_0^2}\right)^{0.86}
+ 0.59\left({k^{2}\over \mu_0^2}\right)
\log\left(2.1\left({k^{2}\over \mu_0^2}\right) +
4.1\right)\right)}\label{axiale}
\end{equation}

As far as the treatment of $\alpha_s$ goes, propagator (\ref{Cl}) uses
a running coupling, which is included in its expression. (We use
a value $\Lambda_{QCD}=0.3$~GeV, as in the perturbative case.) The HKRSW
propagator is at fixed coupling, and renormalization group effects have not
been included. The axial gauge propagator (\ref{axiale}) does include
renormalization group effects. The corresponding $\alpha_s$ is fixed and of
the order of 1 \cite{axial}.
Hence, for propagators (\ref{HKRSW}, \ref{axiale}) we shall use a
fixed $\alpha_s=1$ and indicate how our results scale with $\alpha_s$.

\subsection{Phenomenology of two-gluon exchange}
Each of the above propagators contains an intrinsic scale $\mu_0$: as QCD is a
scale-free theory, this scale cannot be determined directly from
Dyson-Schwinger equations, and must thus be determined through comparison with
some dimensionful quantity. We shall assume from now on that two-gluon exchange
gives the bulk of the cross sections, and that higher orders contribute to the
$s$-dependence of the result. More precisely, as the total cross sections are
well fitted to $22$ mb $s^{\epsilon_0}$ \cite{LD},
we assume that the lowest order must
give us a number of the order of 22~mb. The logarithmic slope at the origin is
a constant of the order of 10~GeV$^{-2}$ plus terms that behave like $\log s$.
Again, we assume that two-gluon exchange will give us the constant piece.
As the
quark-counting rule is essentially independent of energy (at least in the
region where we have data), we assume it must be obeyed too. These are thus
the three tests to which we want to submit the above propagators. Note that it
is not entirely obvious that all three can be passed, as we have three
quantities and one parameter.

We show in Figure~6 the dependence of $\sigma_{tot}^{pp}$ (a), B(0) (b) and
$\sigma_{tot}^{\pi p}/\sigma_{tot}^{pp}$ (c) on the scale $\mu_0$ entering the
propagator. As a growing $\mu_0$ makes the propagator smaller, it is not
surprising that the total cross section goes down with the propagator scale,
as shown in Figure~6(a).
It is less obvious however that for $\mu_0\sim 0.3-1.0$~GeV, one gets a large
suppression factor, and thus each propagator can give a good starting value
for the total cross section, of the order of 20 mb.
The logarithmic slope of the elastic cross section also gets cured by the
introduction of $\mu_0$, and numbers of the order of 10~GeV$^{-2}$ can
be achieved for scales of the same order. Note in passing that if we were to
take these results at face value, propagator~(\ref{Cl}) would have to be
disfavoured as it cannot simultaneously reproduce $B(0)$ and $\sigma_{tot}$.
However, the two-gluon exchange process is certainly a very rough
approximation, hence fitting results to this order is certainly misleading.

If we now turn to Figure~6(c), we see that the quark counting rule, which
can be accounted for by the perturbative calculation, is now slightly modified
by nonperturbative effects. For moderate values of $\mu_0$, we get a
result which is smaller than the perturbative one, and thus in better
agreement with the data. For large $\mu_0$, the ratio goes to $2/3$ as the
${\cal E}_2$ terms are more and more suppressed. Figure~6(d) shows the
ratio of the terms for which ${\cal E}_2$ contributes to those which depend
only on ${\cal E}_1^2$ in formula (\ref{hhscatt}). We see that the
Landshoff-Nachtmann suppression is present to some extent for all propagators,
but that only propagator (\ref{HKRSW}) gives a suppression which ensures
the factorizability of the pomeron coupling.

We thus see at this order that nonperturbative effects are non-negligible.
The inclusion of modified propagators in the calculation provides
appreciable improvements, and the improved order $\alpha_s^2$ constitutes
a good starting point for an expansion
in $\log s$. Let us now see to which extent these improvements carry over to
higher orders.

\section{Gluon exchange to order $\alpha_s^3$}
\subsection{General formalism for three-gluon exchange}
The lowest-order QCD diagrams contributing to $C=-1$ exchange between quarks
are those of Figure~7. By contour integration, we obtain the following for
the amplitudes, before the inclusion of the colour algebra:
\begin{eqnarray}
\amp_{(a)}&=&\amp_{(b)}={-64\pi\alpha_s^3 s\log s}\ [I_R]\nonumber\\
\amp_{(c)}&=&\amp_{(d)}={32\pi\alpha_s^3 s\log s}
\ [I_R-i\ \pi\ I_i]\nonumber\\
\amp_{(e)}&=&\amp_{(f)}={32\pi\alpha_s^3 s\log s}
\ [I_R+i\ \pi\ I_i]
\label{3g}
\end{eqnarray}
with
\beq
I_i= \int [\prod_{j=a,c} {d{\vec k}_j \over \left( 2 \pi \right)^2}]
{ \left(2 \pi \right)^2 \delta^{(2)}
\left(\displaystyle{\sum_{j=a,c}} {\vec k}_j -
\qt\right) \over
{\displaystyle{\prod_{j=a,c}} }
\left( {\vec k}_j^{\  2} + \sigma_{j} \right) }
\label{3gim}
\eeq
$I_R$ is a rather complicated function of the momenta, but, as we are about to
explain, we shall not need it in the following.

The colour algebra involves terms like $Tr(\lambda_a\lambda_b\lambda_c)
Tr(\lambda_{a'}\lambda_{b'}\lambda_{c'})$ with $a'$, $b'$, $c'$ some
combination of $a$, $b$, $c$. Using the fact that
$Tr(\lambda_a\lambda_b\lambda_c)=2 (if_{abc}+d_{abc})$,
we recognize that the
amplitude will contain two terms, one proportional to $f_{abc}f_{abc}$
and one proportional to $d_{abc}d_{abc}$.  When we calculate $\bar q q$
scattering instead of $qq$, the first term flips sign while the second
remains the same.
As antiquarks and quarks couple to gluons with opposite signs,
the first term will contribute to the pomeron, while the second will give the
lowest-order odderon. As the pion has the same number of quarks and
antiquarks, the odderon contribution vanishes, whereas for the proton one
obtains a nonzero contribution.

\subsection{The commuting (odderon) case}
As Eq.~(\ref{3g}) shows, when we add all the diagrams of Figure~7
we obtain zero. This means that there are no leading-log contributions
to the odderon at
this order. As $d_{abc}$ is a commuting object, we can apply the QED formalism
developed in Ref.~\cite{book}. For quark-quark scattering, one easily obtains
the third-order result:
\beq
\amp_{\cal O}= {\left(4\pi\right)^3 \alpha_s^3 \over 3} s
\int [\prod_{j=a,c} { d{\vec k}_j\over \left( 2 \pi \right)^2 }]
{\left( 2 \pi \right)^2 \delta^{(2)}
\left(\displaystyle{\sum_{j=a,c}} {\vec k}_j - \qt\right)
\over {\displaystyle \prod_{j=a,c}} \left( {\vec k}_j^{\ 2} +
\sigma_j \right) }
\eeq
for the sum of the diagrams of Figure~7. The quark propagators transform into
$\theta$-functions ordering the vertices in $x^+$ and $x^-$ space. These theta
functions in turn combine together, and one is left with a $\delta$ function,
which makes the $p^+$ and $p^-$ integrations trivial: at every stage, the
quarks remain on the light-cone.

This means that formula (\ref{eikonal}) can be applied directly. Expanding it
to order $\alpha_s^3$, and carrying out the required colour traces, we obtain
no odderon contribution in $p\pi$ scattering whereas $pp$ scattering gives:
\begin{eqnarray}
\amp_{\cal O}&=&-{10\over 9\pi}\alpha_s^3s\int d\kta d\ktb d\ktc\nonumber\\
&\times&{1\over\kta^{\ 2}+\sigma_a}\ \
{1\over\ktb^{\ 2}+\sigma_b}\ \
{1\over\ktc^{\ 2}+\sigma_c}\ \
\delta^{(2)}(\kta+\ktb+\ktc-\qt)\nonumber\\
&\times& [{\cal E}_1(\kta+\ktb+\ktc)
+2\ {\cal E}_3(\kta,\ktb,\ktc)\nonumber\\
&-&{\cal E}_2(\kta+\ktb,\ktc)
-{\cal E}_2(\ktb+\ktc,\kta)
-{\cal E}_2(\ktc+\kta,\ktb)]^2
\label{hhscatt3}
\end{eqnarray}
with ${\cal E}_1$ and ${\cal E}_2$ the two form factors encountered previously
(\ref{f1}, \ref{f2}), and

\beq {\cal E}_3(\kta,\ktb,\ktc)=\int d{\cal M}\ e^{i
\kta.\r_k+i\ktb.\r_l+i\ktc.\r_m}\label{f3}\eeq
with $k\neq l\neq m$.

The third form factor ${\cal E}_3$ corresponds to diagrams where one gluon gets
attached to each quark. As was the case for two-gluon exchange,
formula (\ref{hhscatt3}) is infrared convergent.
This  is due to the fact that ${\cal E}_3$
reduces to ${\cal E}_2$ when one of its momenta vanishes:
\beq
{\cal E}_3(0,\ktb,\ktc)={\cal E}_2(\ktb,\ktc)
\label{f3ir}
\eeq
and similar conditions when $\ktb\rightarrow 0$, $\ktc\rightarrow 0$.

We shall use only a simple
parametrization of ${\cal E}_3$ in this case, based on the
previous one for ${\cal E}_2$ (\ref{f2simple}):
\beq
{\cal E}_3(\kta,\ktb,\ktc)=
{\cal E}_1\left(\kta^2+\ktb^2+\ktc^2-f(\kta.\ktb+\kta.\ktc+\ktb.
\ktc)\right)\label{f3simple}
\eeq
This form gives rise to the expression (\ref{f2simple}) via the use of
Eq.~(\ref{f3ir}).  We can now see why $f=1$ is the solution for the proton
which
corresponds to $f=2$ in the pion case: when the three quarks are scattered
by the same amount in
transverse space, if they have the same longitudinal momentum, they must
recombine into a proton. This is the same argument as for formula
(\ref{Ltermpi}) in the pion case. Thus
\beq
{\cal E}_3(\vec k,\vec k,\vec k)=1
\label{ltermp}
\eeq

This implies that we have $f=1$ in the proton case, and also the form
(\ref{f2simple}) for ${\cal E}_2$.

As already pointed out \cite{Lterm},
the ${\cal E}_3^2$ term of Eq.~(\ref{hhscatt3})
will dominate the elastic amplitude at high-$t$. We can see this from
formula (\ref{ltermp}): configurations for which the three quarks are
scattered by the same amount will not
be affected by the sharp $t$-dependence coming
from ${\cal E}_1$ or ${\cal E}_2$.

Note however that in this formalism, the
exact high-$|t|$ behaviour of the elastic amplitude depends on the details of
the wavefunction. The terms proportional to ${\cal E}_1^2$ will naturally
reproduce the dimensional counting rule at high t (at least once an infrared
regulation makes them finite). However, for ${\cal E}_2$ and ${\cal E}_3$ can
accommodate a wide range of t-dependences: looking at formulae
(\ref{f2simple}, \ref{f3simple}), we
see that depending on the value of $f$, we can
get behaviours similar to $F_1$ for $f=-2$, or totally flat for $f=1$ (2) in
the proton (resp. pion) case. Furthermore, the simple guesses
(\ref{f2simple}, \ref{f3simple}) were motivated by an infrared argument, which
certainly does not account for the process of reference \cite{Lterm}.
Similarly, the only test for the wavefunction (\ref{psih}) has been the
reproduction of the Dirac form factor $F_1$. Hence it is not surprising that a
reproduction of the Landshoff mechanism will necessitate a closer study of the
wavefunction to pin down the $t$-dependence of these terms. We shall simply
mention here that the $t^{-8}$ behaviour {\it can} be implemented in our
calculation, but it relies on a fine-tuning of the hadronic wave function.

On the other hand,  notice that ${\cal E}_3$ is
present (at this order) only for $C=-1$ contributions. Although the odderon is
dominant at high-$t$, it is suppressed at small $t$, because of the
structure of its form factor ${\cal E}_1-3
{\cal E}_2+2{\cal E}_3$. As can be seen from Table 2,  in
pertubative QCD the odderon contribution to the amplitude at $t=0$ is
about 18\% of the two-gluon exchange amplitude.

Thus at this order, the amplitude is purely real, and contains three form
factors. These are the only three possible form factors that will occur at
any order in $pp$ scattering. We notice that the amplitude takes a form that
involves only propagators, so that we are allowed to replace $1/({\vec
k}^2+\sigma)$ by a nonperturbative estimate $D(k^2)$ in Eq.~(\ref{hhscatt3}),
using formula (\ref{KLehmann}).

\subsection{The anticommuting (pomeron) case}
One has to be a little more careful when one evaluates the contributions of
the diagrams of Figure~7 to the pomeron. As we have seen, these $C=+1$
contributions will come from the anticommuting part of the colour algebra, and
thus will flip the sign of some of
the expressions of Eq.~(\ref{3g}). This means that
the leading-log contributions will not cancel anymore, and thus we shall have
a $\log s$ contribution. A na\"\i ve application of formula (\ref{eikonal}),
taking ordering in the $x^+$ and $x^-$ directions leads to the counterpart of
Eq.~(\ref{hhscatt3}):
\begin{eqnarray}
\amp_{\cal P}&=&{4\over 3\pi^2}n_1
n_2 \alpha_s^3 is\log s\int d\kta d\ktb d\ktc\nonumber\\
&\times&{1\over\kta^{\ 2}+\sigma_a}\
{1\over\ktb^{\ 2}+\sigma_b}\
{1\over\ktc^{\ 2}+\sigma_c}\
\delta^{(2)}(\kta+\ktb+\ktc-\qt)\nonumber\\
&\times& {\cal F_P}(\kta,\ktb,\ktc)
\label{hhscatt3pom}
\end{eqnarray}
with
\begin{eqnarray}
{\cal F_P}(\kta,\ktb,\ktc)&=&
{1\over 2}[{\cal E}_1^{h_1}(\kta+\ktb+\ktc)
-{\cal E}_2^{h_1}(\kta+\ktb,\ktb)]\nonumber\\
&\times& [{\cal E}_1^{h_2}(\kta+\ktb+\ktc)
-{\cal E}_2^{h_2}(\kta+\ktb,\ktc)\nonumber\\
& &
-{\cal E}_2^{h_2}(\ktb+\ktc,\kta)
+{\cal E}_2^{h_2}(\ktc+\kta,\ktb)]\nonumber\\
& &+\{h_1\leftrightarrow h_2\}
\label{wrongf}
\end{eqnarray}

It is interesting that once again, the expressions (\ref{3g}) will combine to
allow the use of nonperturbative propagators, via the integration of the
$1/({\vec k}^2+\sigma)$ with a K\"allen-Lehmann density,
as explained in the two-gluon
case (\ref{KLehmann}).

The $\log s$ comes from large rapidity gaps, {\it
i.e.} one of the gluons will have to carry both a large $+$ and a large $-$
components (by ``large'' we mean bigger than $O(1/s)$). But one must then be
very careful that the final state does not develop a large mass of the order
of $\sqrt{s}$: if a line going in the $+$ direction with momentum $\sqrt{s}$
absorbs a momentum $k_-$ going in the $-$ direction, it will develop a mass
$\sqrt{s} k_->> m_h$, and thus will be able to combine with the other quarks
only after another gluon has carried that $k_-$ component. For example, we
show in Figure~8 a configuration which is always suppressed.

The effect of these large masses will be to destroy some of the contributions
to formula (\ref{hhscatt3pom}). In order to decide which terms get suppressed,
we use the fact that hadron-hadron scattering has the same structure as
photon-photon scattering. Hence, although in the hadronic case the use of
cutting rules is obscured by the fact that some of the propagators are
absorbed in the wavefunction, as illustrated by
Eq.~(\ref{psigg}), the argument of
Ref.~\cite{BFKL} can be applied to the hadronic case.
In both cases, the propagators that
generate a large final-state mass are suppressed, so that the remaining form
factors are the same as those derived in the eikonal approximation. Rederiving
the results of reference \cite{BFKL}, we obtain,
instead of (\ref{wrongf}) in Eq.~(\ref{hhscatt3pom}):

\begin{eqnarray}
{\cal F_P}(\kta,\ktb,\ktc)&=&
{1\over 2}{\cal F}_1^{h_1}{\cal F}_1^{h_2}
 -{1\over 4}({\cal G}_a^{h_1}{\cal G}_a^{h_2}
+{\cal G}_c^{h_1}{\cal G}_c^{h_2})\nonumber\\
&+&{\cal F}_1^{h_1}({\cal G}_a^{h_2}+{\cal G}_c^{h_2})
+{\cal G}_a^{h_1}{\cal G}_c^{h_2}\nonumber\\
& &+(h_1\leftrightarrow h_2)\nonumber
\label{rightf}
\end{eqnarray}
where ${\cal F}_1={\cal E}_1(\kta+\ktb+\ktc)$,
${\cal G}_a={\cal E}_2(\kta,\ktb+\ktc)$ and ${\cal G}_c={\cal E}_2(\ktc,
\kta+\ktb)$.

\subsection{The complete pomeron to order $\alpha_s^3$}
In order to compute the full order $\alpha_s^3$ contributions to the pomeron,
we need to add the H and Y diagrams of Figure~9. As is well known \cite{book},
when the
gluons are exchanged between two quark lines, these diagrams mostly cancel, and
give a contribution proportional to ${\cal E}_1^2$. It
turns out that this almost
exact cancellation holds also for the diagrams where different quarks within
the same hadron contribute. That is because the Y diagram contributes in the
configuration of Figure~10(a), which reproduces exactly the form factor of
Figure~10(b): each line is hit by the same amount.

Although the Soper-Gunion formalism does not explicitly include these, it is
not very hard to convince oneself that the form factors will be the same as in
two-gluon exchange, so that we get for the sum of the diagrams of Figure~7:
\begin{eqnarray}
&-&{4\over 3 \pi^2} \qt^{\ 2} n_1 n_2 \alpha_s^3 i s \log s \int d\kta
d\ktb \nonumber\\
&\times& { \left( {\cal E}_1^{h_1}(\qt) - {\cal E}_2^{h_1}
(\qt-\kta,\kta) \right)
\left( {\cal E}_1^{h_2}(\qt) - {\cal E}_2^{h_2}(\qt-\ktb,\ktb) \right)
\over
(\kta^{\ 2}+\sigma) \left( (\qt - \kta)^{\ 2} + \sigma \right)
(\ktb^{\ 2}+\sigma) \left( (\qt - \ktb)^{\ 2} + \sigma \right) }
\label{HY}
\end{eqnarray}
with ${\cal E}_1$ and ${\cal E}_2$ the form factors introduced in formulae
(\ref{f1},\ref{f2}). Notice that, once more,
assuming the existence of a K\"allen-Lehmann
density for the gluon propagator enables us to change $1/({\vec k}^2+\sigma)$
into $D(k^2)$.

\section{Phenomenology of the order $\alpha_s^3$ results}
\subsection{General structure of the answer}
As we have shown in the previous sections, the complete order $\alpha_s^3$
formalism leads to a hadronic amplitude that has the following form:
\beq
\amp(s,t)= \amp_2\left\{ i\left[1+\log s \left(\epsilon_0+\alpha
't+O(t^2)\right)\right]+f_{odd}(t)\right\}
\label{structure}
\eeq

It is of course tempting to see in this a first-order Taylor expansion in
$\log s$ of a pomeron pole, plus a zeroth-order term from an odderon pole.
There would then be a one-to-one mapping
between the Regge picture of the pomeron and this calculation. As BFKL have
shown \cite{BFKL}, life is not so simple,
and higher-order terms spoil the analogy. Hence in the following
the terms ``pomeron
intercept'' or ``slope'' must not be taken literally.
As we shall now see, all the
problems of the BFKL formalism \cite{RH} are
already present at this low order. It is thus worth examining them in the
context of the simple equations we have gathered in this paper, rather than
obscuring the issues by resumming.

At this order, the normalization of the cross section, $\amp_2$,
comes from two-gluon exchange, formula (\ref{hhscatt}). The pomeron intercept,
$1+\epsilon_0$, is the ratio of two to three gluon exchange, formula
(\ref{hhscatt}, \ref{hhscatt3pom}). The odderon contribution $f_{odd}$
comes of course from
Eq.~(\ref{hhscatt3}). Finally, the $\alpha '$ contribution will come from
Eq.~(\ref{HY}) and from the Taylor expansion of Eq.~(\ref{hhscatt3pom}).

The perturbative results are shown in Table 2.
As we have mentioned earlier, the modifications to the gluon
propagator that we shall discuss in the following will diminish the dependence
of the results on the choice of form factors. Therefore, the $25\%$ variations
that we get in $\epsilon_0$ and $\sigma_2$ can be considered as upper bounds
on the theoretical uncertainty from form factors. For simplicity, we shall
only consider the parametrization (\ref{f1pp}, \ref{f2simple},
\ref{f3simple}) in the following.

The interest of running
$\alpha_s$ becomes obvious at third order. The main problem here is that
$\epsilon_0$ is much too large. Running the coupling will reduce the effect of
the third gluon, and thus lower $\epsilon_0$. As Table 2 shows, it does not
decrease it enough. The running $\alpha_s$ that we used can in fact be
mimicked by a fixed $\alpha_s=0.72$. Similarly, the odderon contribution
is in turn
further suppressed by the running, to about 8\% of the 2-gluon exchange
amplitude.

Another obvious problem
is the fact that, although the
expression (\ref{HY}) is finite for $\qt\rightarrow 0$, it does not lead to a
finite value for $\alpha '$: the integrands at $\qt=0$ look like
$({\cal E}_1-{\cal E}_2)/k^4$, and we thus have
a logarithmically divergent $\alpha '$.

Finally, we show that at this order the pomeron does not appear to be a
universal object: we calculated $J/\psi\ p$ scattering, using a monopole form
factor for the $J/\psi$ similar to Eq.~(\ref{f1ppi}) but with
$\sqrt{<r_{J/\psi}^2>}=0.2$~fm \cite{halzen}. We then obtain a 15\% difference
in the coefficient of the $\log s$ term.

\subsection{Nonperturbative effects}
As we have already mentioned, we can easily accommodate nonperturbative
propagators in the expression given for the hadronic amplitude. Provided that
the propagators have the usual analytic properties
embodied by the K\"allen-Lehmann
density, the prescription is simply to replace $1/(k^2+\sigma)$ by the
nonperturbative prescription for the propagator, $D(k^2)$. Note that the
reason why such a simple prescription works remains obscure, as in principle
a more complicated structure for the hadronic amplitude is allowed. In fact,
were the real part $I_R$ of Eq.~(\ref{3gim}) not
to cancel, the situation would be
more complicated as the integral over $\sigma$ would not simply reproduce the
gluon propagators. It
happens, at least at this order of perturbation theory, that the amplitude
effectively involves only $t$-channel propagators.
Before describing the effect of propagators
(\ref{HKRSW}, \ref{Cornwalle}, \ref{axiale}) on the amplitude,
we wish to point out
that a simple bound will apply at this order, provided the propagator
does not flip sign.

The form factor in the odderon part of the amplitude (\ref{hhscatt3})
provides a more effective suppression than the form factor entering the
pomeron contribution (\ref{hhscatt3pom}). One therefore obtains:
\beq
f_{odd}\leq {54\over 5\pi}\epsilon_0
\label{alphaodd}
\eeq
If some mechanism can bring $\epsilon_0$ to be in agreement with data, and
hence be of the order of 0.1, we expect the lowest-order odderon to be
similarly modified, and hence relation (\ref{alphaodd}) to be still true.

Figure~11 shows in details the result of the use of propagators
(\ref{Cornwalle}, \ref{HKRSW}, \ref{axiale}), as a function of the
nonperturbative scale $\mu_0$. Figure~11(a) shows that $\epsilon_0$
can in principle get as low as 0.35. For
values of $\mu_0$ favoured by our previous
discussion on two-gluon exchange, we see that we get values of the order of 2
for the intercept: it thus seems impossible to get acceptable numbers
both for two and three-gluon exchange. Another way to look at this is to allow
$\alpha_s$ to vary: for values around 0.1, one would indeed get an intercept
compatible with data, but that would mean that although rising at the correct
rate, the pomeron cross section would be much too low.

Figure~11(b) shows that the pomeron slope becomes finite once the infrared
region is smoothed out. Values compatible with 0.25~GeV can again be achieved,
but again for values of $\mu_0$ or $\alpha_s$ that would suppress
two-gluon exchange.

Figure~11(c) shows the pomeron trajectories that we obtain for the
values of $\mu_0$ giving 22~mb for the total two-gluon cross section, {\it
i.e.} 0.282~GeV for (\ref{Cornwalle}), 0.491~GeV for (\ref{HKRSW}) and 0.946
GeV for (\ref{axiale}). The trajectories are
much flatter than in the perturbative case, and that
their slopes are finite.

Finally, we show in Figure~11(d) that the ratio of the odderon amplitude
to the two-gluon exchange one tends to be
greater or equal to the perturbative answer. This simply means that it falls
more slowly with increasing $\mu_0$ than two-gluon exchange. Similarly,
$\epsilon_0(J/\psi\ p)$ is again roughly 10 to 15\% larger than
$\epsilon_0(pp)$.

Hence there is no ``golden propagator".
The three propagators we have tried have
met with some success, but none of the gives us a perfect fit.
None of them can accommodate a sizeable two-gluon exchange amplitude together
with a slowly rising third-order amplitude. At third order, propagator
(\ref{Cornwalle}) gives the most promising results. One has to point out
however that this is mainly due to the fact that the $\alpha_s$ used is
appreciably smaller than in the case of the two other propagators.

\section{Conclusion}
We have shown in this paper that the Soper-Gunion formalism can be extended to
include three-gluon effects, and that the proton possesses three form factors,
in the valence quark approximation. The structure of these form factors
naturally suppresses the odderon at small $t$ and makes it leading at high-$t$.

The preceeding study has also shown that the infrared region is important
in leading $\log s$ calculations. Although the
perturbative answer is infrared finite, the
dominant region of $k^2$ remains small until $t$ is as large as $10$~GeV. We
have shown that nonperturbative effects can change the answer by large
factors, of the order of 3. Before meaningful comparisons can be made with
deep inelastic experiments, or with diffractive ones, one will have to
understand this region better.

The use of recent solutions to the Dyson-Schwinger equation, combined with
the simple idea that it is enough to modify the gluon propagator at small
momentum transfer \cite{LN}, is not sufficient to diminish the value of the
pomeron intercept to bring it in agreement with data.

It is surely possible to find a functional form for the gluon propagator that
will bring the result in closer agreement with experiment. This can in
principle be repeated order by order. Given the wide variety of behaviours at
the origin which we have examined in this paper, the functional form of the
propagator should have one of the following forms: either be suppressed near
the origin faster that propagator (\ref{HKRSW}), or change sign in the
$t$-channel. It is however unlikely that such a
phenomenological procedure would lead to a propagator exhibiting the correct
properties at high $k^2$.

One could of course criticize the use of a loop expansion in a
region which is {\it a priori} nonperturbative. However, it is interesting to
notice that the perturbative calculation gives an answer which is
qualitatively correct. Hence the idea of a loop expansion, or even
the concept of quarks and gluons seem to be fruitful in this context.

Another possibility is that at these high values of $\alpha_s$, the loop
expansion exhibits its asymptotic character very early, and that the
lowest-order term is as far as we will ever get. The use of nonperturbative
propagators in the lowest-order expression has already led to a qualitatively
successful phenomenology \cite{LN,halzen,CR,phenotwo}. It
could be that the third order
already departs from the true answer.

Before jumping to such drastic conclusions, one has to re-examine the
procedure used to derive the results of this paper.

First of all, we are using a leading-$\log s$ procedure. For this to make
sense, the sub-leading terms from the third order must be smaller than the
second order. We can get a rough estimate of the size of these terms: $s$ to
$u$ crossing symmetry of the amplitude implies that $\log s$ must be replaced
by $(\log(s)+\log(-s))/2 = \log s-i\pi/2$. Hence the ratio of the sub-leading
terms to the leading ones of the previous order should be about
$i\pi\epsilon_0/2$. For the magnitude of this to be small, we certainly need
$\epsilon_0<<2/\pi$. Therefore, unless the intercept $\epsilon_0$ is small,
the leading-log $s$ calculation probably does not make sense. For the
perturbative calculation, demanding that sub-leading terms are half of the
leading ones at the preceeding order leads to: $\alpha_s<0.15$. As we are
trying to reduce the value of the intercept, this would at the same time lower
the sub-leading terms and put the leading-log $s$ approximation on firmer
ground.

Two basic ingredients need
to be implemented before one can present final conclusions.
First of all, the simple replacement of the gluon propagator by a solution of
the Dyson-Schwinger equation explicitly breaks gauge invariance in the H and Y
diagram of Figure~10. Explicit solutions for the vertices are known
\cite{BBZ}, and such a procedure would definitely modify our estimate of
$\alpha'$.

Furthermore, the fact that the three-gluon exchange diagrams of
Figure~6 involve only gluon propagators in the $t$-channel remains puzzling,
as the answer can be obtained by cutting rules, and as some of the cuts involve
the gluons. There must thus be a precise interplay between the gluon and quark
propagators for this property to be conserved. A re-examination of the cutting
rules in the context of modified propagators seems needed. Indeed, for
three-gluon exchange, the cuts through quark lines contribute precisely $-1/2$
of the cuts though gluon lines. Were the cutting rules to be modified by
nonperturbative effects, this delicate balance could be broken and the result
turn out to be entirely different. We plan to address these questions in the
future.
\bigskip

\noindent {\Large {\bf Acknowledgements}} \\
We wish to acknowledge discussions with
P.V. Landshoff, L.N. Lipatov, B. Margolis and D.~Soper.
This work was supported in part by NSERC (Canada) and les fonds FCAR
(Qu\'ebec).

\newpage
\begin{center}
\begin{tabular}{||c|c|c||} \hline \hline
 quantity                 &fixed $\alpha_s$ &running $\alpha_s$ \\ \hline
$\sigma_2(pp)/\alpha_s^2$           & 75.8 (59.6) mb & 63.5 (46.6) mb\\ \hline
 $B(0)$                             &$\infty$ & $\infty$ \\ \hline
 $\sigma_2(p\pi)/\sigma_2(pp)$ & 0.654   & 0.660   \\ \hline
\end{tabular}

Table 1: the perturbative results at order $\alpha_s^2$
\end{center}
\bigskip
\begin{center}
\begin{tabular}{||c|c|c||} \hline \hline
 quantity                 &fixed $\alpha_s$ &running $\alpha_s$ \\ \hline
 $\alpha_0 (pp)/\alpha_s$           & 1.87 (1.82)   & 1.36 (1.20)   \\ \hline
 $\alpha_0(p\pi)/\alpha_s$          & 1.75    & 1.25 \\ \hline
 $\sigma_{odd}/\alpha_s\sigma_2$    & 0.174   & 0.079    \\ \hline
 $\alpha '$                         &$\infty$ & $\infty$\\ \hline
 $\alpha_0 (J/\psi p)/\alpha_0 (pp)$ &1.15& 1.23 \\\hline\hline
\end{tabular}

Table 2: the perturbative results at order $\alpha_s^3$
\end{center}
\newpage
\begin{center}
{\bf FIGURE CAPTIONS}
\end{center}
Figure 1: Two-gluon exchange contributions to quark-quark
scattering. The dashed lines represent gluons and the plain ones represent
quarks.

\nnn Figure 2: Two-gluon exchange contributions to photon-photon
scattering. (a) gives rise to the ${\cal E}_1(\kta+\ktb)$ and (b) to
${\cal E}_2(\kta,\ktb)$.

\nnn Figure 3: The Dirac form factor ${\cal E}_1$ from formula (2.12)
(dashed curve), compared with the form resulting from a dipole parametrization
of $G_E$ \cite{ppform} (plain curve).

\nnn Figure 4: The elastic differential cross section at
$O(\alpha_s^4)$, as predicted by perturbative QCD. The plain curve is
for fixed $\alpha_s=1$ and scales like $\alpha_s^2$,
the dashed one for a running $\alpha_s$ (see text).

\nnn Figure 5: The average gluon momentum flowing through the perturbative
two-gluon exchange graphs for fixed $\alpha_s$ (plain curve) and running
$\alpha_s$ (dot-dashed curve).

\nnn Figure 6: Two-gluon exchange results for
three nonperturbative gluon propagators,
Cornwall (plain curve), Eq.~(\ref{Cornwalle}),
H\"abel-K\"onnig-Reusch-Stingl-Wigard
(dashed curve),\break Eq.~(\ref{HKRSW}) and Cudell-Ross (dot-dashed curve),
Eq.~(\ref{axiale}). The horizontal axis gives the nonperturbative scale
entering the propagator. (a) shows the total proton-proton cross section,
which scales like $\alpha_s^2$, (b) gives
the logarithmic slope at $t=0$ of the elastic cross section, (c)
the ratio of the pion-proton to the proton-proton cross sections and (d) the
ratio of the terms in the amplitude where gluons hit only one quark line to
those where gluons hit two quark lines.

\nnn Figure 7: Three-gluon exchange contributions to hadron-hadron
scattering. We show the diagrams proportional to
${\cal E}_1(\kta+\ktb+\ktc)^2$.

\nnn Figure 8: One of the diagrams that always gets suppressed because the
final state develops a large mass.

\nnn Figure 9: Some of the H, X and Y diagrams contributing to the
pomeron slope.

\nnn Figure 10: Both diagrams give rise to the same form factor
${\cal E}_2(\kta,\qt-\kta) \ {\cal E}_1(\qt)$.

\nnn Figure 11: Order $\alpha_s^3$ results
for the three nonperturbative gluon propagators
of Figure 6. Our convention for the curves is the same as in Figure 6. $\mu_0$
is the nonperturbative scale entering the propagator.
 All graphs scale like $\alpha_s$.
(a) shows the ratio at $t=0$ of the
coefficient of the $\log s$ term to the two-gluon
exchange term; (b) the ratio at $t=0$ of the coefficient of the $t \log s$
term to the two-gluon exchange term; (c) shows the pomeron trajectory at this
order for the values of $\mu_0$ which give 22~mb for the two-gluon exchange
cross section, the thick curve is the trajectory in the perturbative answer;
 (d) gives the ratio of the third order odderon amplitude to
the imaginary part of the two-gluon amplitude.



\newpage

\begin{thebibliography}{99}
\bibitem{LD} A. Donnachie and P.V. Landshoff,
\ Nucl. Phys. {\bf B244} (1984)
 322; {\bf B267} (1985) 690; Phys. Lett. {\bf B296} (1992) 227
\bibitem{BFKL}E.A. Kuraev, L.N. Lipatov and V.S. Fadin, \ {\it Sov. Phys.
JETP} {\bf 45} (1977) 199\\
 Y.Y. Balitski\v \i\ and L.N. Lipatov, \ {\it Sov. J. Nucl. Phys.}
 {\bf 28} (1978) 822
\bibitem{goulianos} K. Goulianos, Physics Reports {\bf 101} (1983) 169
\bibitem{Richards}D.G. Richards, Nucl.Phys. {\bf B258} (1985) 267
\bibitem{Lterm}A. Donnachie and P.V. Landshoff, Phys. Lett. {\bf 123B} (1983)
345, P.V. Landshoff, Phys. Rev. {\bf D 1024} (1974) 1024
\bibitem{RH}R.E. Hancock and D.A. Ross, Nucl.Phys. {\bf B394} (1993) 200
and {\bf B383} (1992) 575
\bibitem{LN} P.V. Landshoff and O. Nachtmann, \ Z. Phys. {\bf C35} (1987) 405
\bibitem{axial} J. R. Cudell and D.A. Ross, \ Nucl. Phys. {\bf B359} (1991)
247\\
J.R. Cudell, \ Proceedings of the $4^{th}$ Blois Workshop on Elastic \&
Diffractive Scattering, La Biodola, Italy (1991).
\bibitem{feynman}U. H\"abel, R. K\"onning, H.G. Reusch, M. Stingl and S.
Wigard,
preprint Print-89-0128 (MUNSTER) and Z. Phys. {\bf A336} (1990) 435
\bibitem{Cornwall}J.M. Cornwall, Phys.Rev. {\bf D26} (1982) 1453
\bibitem{Low}F.E. Low, Phys. Rev. {\bf D12} (1975) 163; S. Nussinov, Phys.
Rev. Lett. {\bf 34} (1976) 1286
\bibitem{photon} H. Cheng and T.T. Wu, \ Phys. Rev. Lett. {\bf 24} (1970)
1456\\
V.N.Gribov, L.N. Lipatov and G.V. Frolov,
\ Sov. J. Nucl. Phys. {\bf 12} (1971) 543
\bibitem{GS}J.F. Gunion and D. Soper, Phys. Rev. {\bf D 15} (1977) 2617
\bibitem{ppform} P.N. Kirk et al., Phys. Rev. {\bf D8} (1973) 63\\
 R.H. Hodstadter, F. Bumiller and M.R. Yearian,
Rev. Mod. Phys., {\bf 30} (1958) 482
\bibitem{piform}NA7 collaboration, S.R. Amendolia
et al., Nucl. Phys. {\bf B277} (1986) 168
\bibitem{halzen} M.B. Gay Ducati, F. Halzen and A.A.Natale,
preprint MAD-PH-750(1993)\\
 F. Halzen, G.I. Krein and A.A. Natale, Phys. Rev. {\bf D47} (1992) 295
\bibitem{CR}J.R. Cudell and D.A. Ross, preprint McGill 92-50
\bibitem{book} H.C. Cheng and T.T. Wu,
{\it Expanding Protons: Scattering at
High Energies} (MIT Press: Cambridge, 1987)
\bibitem{lattice}J.E. Mandula and M. Ogilvie,
Phys. Lett. {\bf B185} (1987) 127\\
 P.A. Amundsen and J. Greensite,
Phys.Lett. {\bf B173} (1986) 179\\
 S.P. Li, Phys. Rev. {\bf D32} (1985) 3321
\bibitem{Zwanziger}D. Zwanziger, Nucl. Phys. {\bf B323} (1989) 513
\bibitem{bao} B.U. Nguyen, Ph. D. thesis (1993)
\bibitem{BBZ}M. Baker, J.S. Ball and F. Zachariasen, Nucl. Phys. {\bf B186}
(1981) 531 and references therein
\bibitem{phenotwo}J.R. Cudell, Nucl. Phys. {\bf B336} (1990) 1\\
  A. Donnachie
and P.V. Landshoff, Nucl. Phys. {\bf B311} (1989) 509
\end{thebibliography}
\end{document}